\documentclass[a4paper, reqno, 10pt]{amsart}

\usepackage[utf8]{inputenc} 
\usepackage{amsmath, amssymb, amsthm}
\usepackage{mathrsfs}
\usepackage{mathtools}
\usepackage{dsfont}
\usepackage{array}
\usepackage{multirow}
\usepackage{booktabs}
\usepackage{microtype}
\usepackage{subcaption}
\usepackage{pgfplots}
\usepackage{url}
\usepackage{bm}

\usepackage{algorithm}
\usepackage[noEnd=true]{algpseudocodex}
\usepackage{caption}
\algnewcommand{\Compute}{\State \textbf{compute} }
\algnewcommand{\Precompute}{\State \textbf{precompute} }
\newcommand\hh[1]{\vphantom{\rule{0pt}{#1}}}

\usepackage{listings}

\usepackage[giveninits, maxbibnames=99, backend=biber, eprint=false, url=false, doi=true, isbn=false, sorting=nyt, sortcites=true]{biblatex}
\usepackage[hyperfootnotes=false]{hyperref}

\graphicspath{{figures/}}

\pgfplotsset{compat=newest}
\usetikzlibrary{arrows.meta, shapes, fit}

\newcommand\tikzgraphsettings{\tikzset{
	every node/.style = {
		circle, fill,
		inner sep=1pt,
		minimum size=1.7mm,
		outer sep=1pt,
		font=\footnotesize,
		text=white,
	},
	every label/.append style={font=\footnotesize, text=black},
	every mark/.append style={
		mark size=2.25pt
	},
	empty/.style = {
		font = \small,
		rectangle,
		inner sep=0pt,
		draw=none,
		fill=none
	},
	>=latex}
}
    
\tikzset{
  contour/.style={
    decoration={
      name=contour lineto closed,
      contour distance=#1
    },
    decorate}}

\definecolor{mplblue}{HTML}{1f77b4}
\definecolor{mplorange}{HTML}{ff7f0e}
\definecolor{mplgreen}{HTML}{2ca02c}
\definecolor{mplred}{HTML}{d62728}
\definecolor{mplpurple}{HTML}{9467bd}
\definecolor{mplbrown}{HTML}{8c564b}
\definecolor{mplpink}{HTML}{e377c2}
\definecolor{mplgray}{HTML}{7f7f7f}
\definecolor{mplyellowgreen}{HTML}{bcbd22}
\definecolor{mplcyan}{HTML}{17becf}

\theoremstyle{plain}
\newtheorem{theorem}{Theorem}
\newtheorem*{theorem*}{Theorem}

\theoremstyle{definition}
\newtheorem{definition}{Definition}
\newtheorem{example}{Example}

\theoremstyle{remark}
\newtheorem{remark}{Remark}

\DeclareMathOperator*{\argmax}{argmax}

\DeclareMathOperator{\pc}{PC}
\DeclareMathOperator{\nc}{NC}
\DeclareMathOperator{\diam}{diam}

\newcommand{\sh}{\phi^{\mathrm{Sh}}}

\newcommand\restr[2]{{\left.\kern-f lldelimiterspace#1\mathchoice{\vphantom{\big|}}{}{}{}\right|_{#2}}}

\title[PAGTC for set-function maximization on networks]{Past-aware game-theoretic centrality:\\ a framework for cardinality-constrained set-function maximization on networks}

\author{Francesco Zigliotto}
\address{Scuola Normale Superiore. Pisa, PI 56126, Italy.}
\email[Corresponding author]{francesco.zigliotto@sns.it}


\usepackage{silence}
\begin{document}

\begin{abstract}
\looseness-1 We consider cardinality-constrained optimization of set functions over the nodes of a graph. The standard greedy algorithm selects each node according to its immediate marginal contribution, a local criterion that may fail to anticipate the synergies within the final set. We introduce \emph{past-aware game-theoretic centrality} (PAGTC), which evaluates a candidate node through its expected marginal contribution over the possible completions of the current partial solution to the prescribed target size. This yields a sequential selection strategy that explicitly accounts for the final budget. For nonnegative monotone submodular objectives and a budget~$r$, we prove an approximation guarantee of $r/(2r-1)$ and derive computable \emph{a posteriori} bounds. Since direct PAGTC evaluation involves averaging over a large number of coalitions, we extend an exact computation framework for game-theoretic centrality and derive efficiently computable expressions for two classes of graph-optimization problems, namely facility location and influence in complex contagion, covering both submodular and non-submodular cases. The numerical results show that the benefits depend on the objective and are most pronounced for complex contagion, where submodularity does not hold.
\end{abstract}

\maketitle
\thispagestyle{empty}

\section{Introduction}

Many network problems require selecting a fixed-size subset of nodes that act collectively. Classic examples include locating facilities~\cite{NemhauserBank}, identifying dense subgraphs~\cite{DensestSubgraphReview}, and selecting seeds for a contagion process~\cite{CCReviewCentola}. These tasks naturally reduce to maximizing a set function $f:2^V\to\mathbb{R}$, which quantifies the joint value of any subset of the graph's node set~$V$. The standard baseline for solving this cardinality-constrained maximization is the \emph{greedy} approach, which iteratively builds a solution by adding the node with the highest immediate marginal contribution. Several algorithms have been built upon this strategy, particularly when the objective set function~$f$ is \emph{submodular}~\cite{AcceleratedGreedy, Krause_Golovin_2014}, meaning that the marginal benefit of adding a new element diminishes as the underlying set grows. For monotone submodular functions, the standard greedy approach provides a strong theoretical performance guarantee~\cite{NemhauserOptimality,nemhauser1978}. In the inherently more complex non-submodular case, however, evaluations based solely on immediate marginal gains are more prone to miss synergies that become relevant only within the final set. Researchers have established weaker bounds for the standard greedy algorithm by leveraging parameters like the \emph{submodularity ratio}, which quantifies a function's deviation from submodularity~\cite{KnapsackConstraint}, alongside proposing improvements to the solution via iterative node exchanges~\cite{Yang_2020}.

Centrality measures provide a natural way to evaluate the role of individual nodes in such collective problems. Specifically, \emph{game-theoretic centrality} borrows concepts from cooperative game theory~\cite{ShapleyValue, Semivalues} and evaluates the importance of a node through its expected marginal contribution to coalitions of other nodes. Both the theoretical properties~\cite{AxiomaticCharacterization} and the algorithmic aspects~\cite{BetweennessGameTheoretic, CalculationShapley} of these measures have been investigated, and an overview of the main approaches is provided in~\cite{GTCReview}. 

Standard game-theoretic centrality, however, evaluates candidates over a random distribution of coalitions, treating all potential collaborators as uncertain. This is ill-suited for a sequential node-selection process, in which a partial solution has already been fixed and only its completion remains unknown. At the other extreme, the standard greedy algorithm accounts exactly for the nodes already selected, but limits the evaluation of each candidate to its immediate marginal contribution. To combine these two perspectives, we introduce the \emph{past-aware game-theoretic centrality} (PAGTC), a variant of game-theoretic centrality in which the expected contribution of a node is conditioned on the presence of a predetermined subset of collaborators (the \emph{past}). 

\looseness-1
In our approach, we leverage PAGTC scores to construct a solution iteratively. At each step, a candidate node is evaluated through its expected contribution to a final group of the prescribed size, conditioned on the nodes already selected. Thus, unlike the standard greedy heuristic, the proposed criterion also accounts for the possible completions of the current partial solution. The scores are dynamically updated after every selection, so that uncertainty about the final group progressively decreases. 

\looseness-1
The practical viability of the framework depends on whether the PAGTC scores associated with the objective function~$f$ can be computed efficiently. Their direct evaluation requires averaging over a large number of coalitions and would therefore be impractical within an iterative algorithm. Building on existing computational techniques for game-theoretic centrality~\cite{BetweennessGameTheoretic, CalculationShapley}, we extend an efficient exact-computation framework to the past-aware setting.

We then apply it to two classes of graph-optimization problems, namely facility location and influence maximization, covering both submodular and non-submodular cases. For each class, we derive exact formulas that avoid coalition enumeration and lead to algorithms with the same time complexity as the greedy methods. Thus, our benchmarks focus on the greedy baseline to highlight the effect of target-size awareness, rather than comparing against specialized state-of-the-art algorithms.

We find that the PAGTC approach proves particularly effective for \emph{influence maximization} under \emph{complex contagion} dynamics. This model describes the spread of a contagion or information over a network when activation requires reinforcement from multiple neighbors~\cite{centola2007, CCReviewCentola}. The corresponding influence functions are non-submodular, and the standard greedy algorithm no longer retains its classical guarantee. For one-round influence maximization, the derived PAGTC formula yields an algorithm with complexity $O(r|E|)$, where $|E|$ denotes the number of edges, and $r$ is the cardinality of the solution, matching the asymptotic complexity of the corresponding greedy method while producing consistently better solutions in our experiments.

The full-cascade influence-maximization problem is more demanding, since computing PAGTC directly for the complete contagion dynamics is not tractable within our framework. We therefore use the solution obtained for the one-round objective as a candidate for the full-cascade problem. This procedure retains a complexity of $\mathcal O(r|E|)$, making it suitable for large-scale graphs, whereas the greedy algorithm based on repeated simulations of the complete cascade requires $\mathcal O(r|V||E|)$ operations, where $|V|$ is the number of nodes. The experimental results show that the PAGTC candidate is often competitive with the full-cascade greedy solution and becomes particularly effective for stronger reinforcement requirements.

The paper is organized as follows. After introducing the context and notation in Section~\ref{s:notation}, we recall the definition of game-theoretic centrality and present its novel past-aware extension (Section~\ref{s:pagtc}), together with a computational framework (Section~\ref{s:computation_pagtc}). Building on this, we propose a general heuristic for set-function maximization (Section~\ref{s:theory_max}), and provide a theoretical analysis for nonnegative monotone submodular objectives (Section~\ref{s:optimality_bounds}). In Section~\ref{s:threee_examples} we apply our method to two different optimization problems in graphs, deriving
efficiently computable expressions for the corresponding PAGTC score. The resulting algorithms are experimentally evaluated in Section~\ref{s:experiments}. Finally, in Section~\ref{s:conclusions}, we draw conclusions and suggest possible future developments.

\subsection{Notation}
\label{s:notation}

\looseness-1
A \emph{simple undirected graph} $G$ is a pair $(V,E)$, where $V$ is a finite set of nodes and~$E$ is the set of edges. Each edge has the form $\{u,v\}$, with $u,v\in V$ and $u\ne v$. We denote by $N(u)$ the \emph{neighborhood} of node $u$, i.e., the set of nodes $v\in V$ such that $\{u,v\}\in E$. The \emph{degree} of $u$, i.e., the number of its neighbors, is denoted by $\deg(u)$. We say that~$G$ is \emph{connected} if, for any two distinct nodes $u,v\in V$, there exists a \emph{walk} from $u$ to $v$, i.e., a sequence of nodes $w_1,w_2,\dots, w_k$ such that $w_1=u$, $w_k=v$ and  $\{w_i,w_{i+1}\}\in E$ for all $i=1,\dots,k-1$. In a connected graph, the \emph{distance} $d_{uv}$ between two nodes $u$ and $v$ is the number of edges in a shortest walk from $u$ to $v$. We define the \emph{diameter} $\diam(G)$ as the farthest distance between any two nodes. For a group of nodes $S\subseteq V$, we define $d(S, v) = \min_{u\in S} d_{uv}$, with the convention that $d(\varnothing, u)=\infty$. Although extensions to disconnected or directed graphs are possible, for simplicity we assume throughout the paper that~$G$ is a connected simple undirected graph.

We also introduce a useful notation for set-functions. Given a graph $G=(V,E)$, let $f$ be a set function $f:2^V \to \mathbb{R}$, where $2^V$ is the power set of $V$. Then, for any node $u$ and subset $S\subseteq V$, we define the \emph{marginal contribution} of $u$ to $S$ as
\begin{equation}
\label{e:marginal_contribution}
f(u|S)=f\bigl(S\cup\{u\}\bigr)-f(S).
\end{equation}

\section{Past-Aware Game-Theoretic Centrality}
\label{s:pagtc}

This section introduces the main tool of this paper: the past-aware game-theoretic centrality. We begin with a brief review of standard game-theoretic centrality~\cite{GTCReview}.

\subsection{Probabilistic interpretation of GTC}
\label{s:prob_gtc}

Game-theoretic centrality (GTC) indices measure the \emph{collaborative importance} of each node of a graph $G=(V,E)$. They are defined by combining two fundamental components: a (graph-theoretic) \emph{group centrality} and a (game-theoretic) \emph{solution concept}. 

The concept of \emph{group centrality} was introduced in~\cite{GroupCentrality} in order to measure the joint centrality of a set of nodes in a more insightful way than just summing up the single centralities of the nodes in the group. A group centrality $f :2^V\to \mathbb{R}$ assigns a (typically non-negative) real value to every subset of nodes $S\subseteq V$, which quantifies the \emph{collective importance} of the nodes of $S$. Naturally, there are many ways in which such importance can be defined, depending on context. 

To build a game-theoretic centrality, we need to translate a group centrality~$f $ into a node-level centrality, which preserves part of the collaborative information encoded in~$f $. This is achieved by means of a \emph{solution concept}, a tool from cooperative Game Theory. Intuitively, a solution concept assigns a payoff to each player of a collaborative game, where each coalition of players is associated with a value. In our context, the graph nodes are the players, while coalition values are given by~$f $. In principle, a solution concept can be any mapping that associates the group valuation~$f $ with an individual-node centrality. Here we focus on \emph{semivalues}~\cite{Semivalues}, which generalize the \emph{Shapley value}~\cite{ShapleyValue}, one of the most widely adopted solution concepts.

\begin{definition}
\label{d:gtc}
Let $G=(V,E)$ be a graph with $n$ nodes, and let $f $ be a group centrality. The \emph{semivalue game-theoretic centrality} (GTC) of a node $u$ is defined as
\begin{equation}
\label{e:gtc}
\phi^\beta_f (u)=\sum_{s=0}^{n-1}\dfrac{\beta(s)}{\binom{n-1}{s}}\sum_{\substack{S\subseteq V\setminus\{u\}\\|S|=s}}
f (u|S),    
\end{equation}
where $f (u|S)$ is the \emph{marginal contribution} of node $u$ to coalition $S$, defined in~\eqref{e:marginal_contribution} and~$\beta$ is a probability distribution over the integers from $0$ to $n-1$, i.e., a nonnegative function such that
\[
\sum_{s=0}^{n-1}\beta(s)=1.
\]
\end{definition}
The semivalue GTC of a node $u$ can thus be interpreted as the expected marginal contribution of $u$:
\[
\phi^\beta_f (u) = \mathbb E\bigl[f (u|\bm{S})\bigr],
\]
where $\bm S$ is a random subset of $V\setminus\{u\}$ sampled uniformly among the subsets of cardinality $s$, and $s$ is chosen according to $\beta$:
\begin{equation}
\label{e:s_distribution}
\mathbb{P}[\bm S=S] = 
\dfrac{\beta(|S|)}{\binom{n-1}{|S|}}.    
\end{equation}
In this interpretation, $\beta(s)$ represents the probability that $u$ collaborates with a coalition of size $s$. When no prior information is available about the number of collaborators of node~$u$, we can take $\beta$ to be the uniform distribution. The resulting semivalue is called \emph{Shapley value}, and has been characterized axiomatically in~\cite{ShapleyValue}. We denote the corresponding game-theoretic centrality by $\sh_f $.

\begin{example}
\label{e:hospitals}
Consider a graph $G$ where the edges represent the streets of a city, and the nodes represent the crossroads. Assume that a set of facilities of the same type (e.g., schools or hospitals) is to be built. The goal is to place these facilities at the most strategic nodes of the graph. Let a group centrality $f(S)$ quantify the “goodness” of a set $S$ of facilities. For example, if we want to ensure that every location in the city is close to at least one facility, we might employ the \emph{group harmonic centrality}~\cite{BoldiVignaAxiomsCentrality}, defined as
\begin{equation}
\label{e:group_harmonic}
f (S) = \sum_{v\in V\setminus S} \frac{1}{d(S,v)}.    
\end{equation}

\looseness-1
Now, suppose that we need to build a new facility, given that no facilities have been built yet and without any information about the number or positions of future facilities. A natural choice is to select the node $u$ that maximizes the expected value of the marginal contribution $f (u|\bm S)$, computed with respect to a random set~$\bm S$ of other future facilities. This corresponds to choosing the node with maximal game-theoretic centrality $\sh_f (u)$. The computational aspects of evaluating the Shapley value in the particular case where $f $ is the group harmonic centrality are discussed in~\cite{CalculationShapley}.
\end{example}

\subsection{Definition of PAGTC}
\label{s:def}
In this section we introduce the novel concept of \emph{past-aware game-theoretic centrality} (PAGTC). As motivation, consider the scenario described in Example~\ref{e:hospitals}, but with a twist: now we assume that some facilities have \emph{already been built} on a set of nodes $P$, and we need to choose the node $u$ on which to build the next facility. Here, the most natural choice would be to maximize the expected marginal contribution of $u$, conditioned on the fact that $u$ will always collaborate with the nodes in $P$. This motivates the following definition.
\begin{definition}
\label{d:pagtc}
Given a graph $G=(V,E)$, a set of nodes $P$, and a group centrality~$f$, assuming the support of $\beta$ contains at least one size $s \ge |P|$, we define the \emph{semivalue past-aware game-theoretic centrality} of a node $u\notin P$ as
\[
\phi^\beta_f (u|P) = \mathbb E\bigl[f (u|\bm S)\mid \bm S\supseteq P],
\]
where $\bm S$ is a random subset of $V\setminus\{u\}$ with distribution~\eqref{e:s_distribution}, as in GTC, and $\mathbb{E}[\,\cdot\,|\,\cdot\,]$ denotes the conditional expectation. 
\end{definition}

In the context of facility building (Example~\ref{e:hospitals}), $\phi^\beta_f (u|P)$ captures information about both the \emph{future} of the dynamics, by taking the expected value over the future collaborations of $u$, and the \emph{past} of the dynamics, by enforcing collaboration with facilities that have already been built.

We can write PAGTC as in~\eqref{e:gtc}:
\begin{equation}
\label{e:pagtc}
\begin{split}
\phi^\beta_f (u|P)
&=\sum_{S\subseteq V\setminus\{u\}}\dfrac{\mathbb P(\bm S=S,\bm S\supseteq P )}{\mathbb{P}(\bm S\supseteq P)}f (u|S)\\
&=C_\beta\cdot\sum_{s=|P|}^{n-1}\dfrac{\beta(s)}{\binom{n-1}{s}}\sum_{\substack{S\subseteq V\setminus\{u\}\\|S|=s,\,S\supseteq P}}f (u|S),
\end{split}
\end{equation}
where $C_\beta$ is a normalization constant that does not depend on $u$:
\begin{equation}
\label{e:c_beta}
C_\beta=1/\mathbb{P}[\bm S\supseteq P]=\left(\sum_{s=|P|}^{n-1}\beta(s)\dfrac{\binom{n-1-|P|}{s-|P|}}{\binom{n-1}{s}}\right)^{\!\!-1}\!\!.
\end{equation}

\begin{example}
\label{e:beta_concentrated}
If the support of $\beta(s)$ is concentrated at $s=|P|$ (i.e., $\beta=\delta_{|P|}$, the Dirac mass at $|P|$), then $C_\beta=\binom{n-1}{|P|}$ and
\[
\phi^\beta_f (u|P) = f (u|P).
\]
In this case, the PAGTC of a node $u$ is simply the marginal contribution of $u$ to $P$.
\end{example}

\subsection{Closed-form expressions for PAGTC}
\label{s:computation_pagtc}
The computation of (classic) game-theoretic centrality via~\eqref{e:gtc} is computationally infeasible, and Monte Carlo estimations are typically employed~\cite{ShapleyMonteCarlo}. However, a framework was introduced in~\cite{tarkowski2018} which allows to derive closed-form expressions for the semivalue and Shapley GTC that can be computed in polynomial time~\cite{BetweennessGameTheoretic,CalculationShapley,tarkowski2018}. This framework is applicable provided we can express the group centrality $f $ in a specific form, namely:
\begin{equation}
\label{e:efficient_nu}f (S) = \sum_{\substack{\theta \in \Theta \\ \theta \sim S}} g(\theta),
\end{equation}
for a suitable set of “items” $\Theta$, independent of $S$, and a function $g: \Theta \to \mathbb{R}$. Here,~$\sim$ denotes a binary relation between the items in $\Theta$ and the subsets of $V$.

\begin{example}
\label{e:example_int_edges}
One simple example is given by the group centrality
\[
f_{\mathrm{den}}(S)
= \bigl|\bigl\{\{u,v\}\in E: u,v\in S\bigr\}\bigr|
\]
which counts the number of edges of $G=(V,E)$ with both ends in $S$. 
In this case, we can choose $\Theta=E$ and $g\equiv 1$, obtaining
\[
f_{\mathrm{den}}(S) = \sum_{\substack{e\in E\\ e\sim S}}1,
\]
where $e=\{u,v\}\sim S$ whenever $u,v\in S$.    
\end{example}

The framework requires only minimal modifications to compute the newly introduced past-aware version of game-theoretic centrality. Specifically, if $f $ takes the form of~\eqref{e:efficient_nu}, then the following result generalizes \cite[Eq.~(7)]{tarkowski2018}.
\begin{theorem}
\label{t:efficient_semivalue}
Given a graph~$G=(V,E)$, a set of nodes~$P$, and a group centrality~$f $ of the form~\eqref{e:efficient_nu}, the semivalue PAGTC of~$u\in V\setminus P$ can be computed as follows:
\begin{equation}
\label{e:efficient_semivalue}
\phi^\beta_f (u|P)=
C_\beta\sum_{s=0}^{n-1}\dfrac{\beta(s)}{\binom{n-1}{s}}
\sum_{\theta\in\Theta}
g(\theta)\left(\bigl|\pc^s_u(\theta)\bigr|-\bigl|\nc^s_u(\theta)\bigr|\right),
\end{equation}
where $C_\beta$ is given by~\eqref{e:c_beta} and
\begin{equation}
\label{e:pc_and_nc}
\begin{aligned}
\pc^s_u(\theta) &= \{S\subseteq V\setminus\{u\}: |S|=s,\,P\subseteq S,\,\theta\sim S\cup\{u\},\,\theta\not \sim S\}\\
\nc^s_u(\theta) &= \{S\subseteq V\setminus\{u\}: |S|=s,\,P\subseteq S,\,\theta\not \sim S\cup\{u\},\,\theta \sim S\}
\end{aligned}
\end{equation}
denote the sets of \emph{positive} and \emph{negative contributions}.
\end{theorem}
\begin{proof}
Thanks to~\eqref{e:marginal_contribution} and~\eqref{e:efficient_nu}, for any $s=0,\dots,n-1$, we can write
\[
\begin{split}
\sum_{\substack{S\subseteq V\setminus\{u\}\\|S|=s,\,S\supseteq P}}f (u|S)
&=\sum_{\substack{S\subseteq V\setminus\{u\}\\|S|=s,\,S\supseteq P}}\left(\sum_{\substack{\theta\in\Theta\\ \theta\sim S\cup\{u\}}}g(\theta)-\sum_{\substack{\theta\in\Theta\\ \theta\sim S}}g(\theta)\right)\\
& = \sum_{\theta\in\Theta}g(\theta)\left(\bigl|\pc^s_u(\theta)\bigr|-\bigl|\nc^s_u(\theta)\bigr|\right).
\end{split}
\]
The conclusion follows from~\eqref{e:pagtc}.
\end{proof}

\begin{example}
\label{e:example_int_edges2}
For the group centrality $f_{\mathrm{den}}$ of Example~\ref{e:example_int_edges}, the set of negative contributions is always empty, as the internal edges of $S$ never decrease when adding an external node $u$.
As for positive contributions, given an edge $e$, we have that a group of nodes $S\supseteq P$ of size $s$ belongs to $\pc^s_u(e)$ if and only if $e$ is an internal edge of $S\cup \{u\}$ but not of $S$. This can only happen if $e=\{u,v\}$, with $v\in S$. The choices for $S$ are then
\[
\bigl|\pc^s_u(e)\bigr| =
\begin{cases}
\binom{n-1-|P|}{s-|P|} & \text{if $v\in P$} \\[2ex]
\binom{n-2-|P|}{s-|P|-1} & \text{if $v\notin P$}
\end{cases}
\]
which, thanks to Theorem~\ref{t:efficient_semivalue}, yields the following expression:
\[
\begin{split}
\phi_{f_{\mathrm{den}}}^\beta(u|P)
&= C_\beta\sum_{s=|P|}^{n-1}\frac{\beta(s)}{\binom{n-1}{s}}
\left[
    \sum_{\substack{v\in N(u)\\ v\in P}} \binom{n-1-|P|}{s-|P|}+
    \sum_{\substack{v\in N(u)\\ v\notin P}} \binom{n-2-|P|}{s-|P|-1}
\right]\\
&= C^{(1)}_\beta\bigl|N(u)\cap P\bigr| +  C^{(2)}_\beta\bigl|N(u)\setminus P\bigr|,
\end{split}
\]
where
\[
C^{(1)}_\beta = C_\beta\underbrace{\sum_{s=0}^{n-1}\frac{\beta(s)\binom{n-1-|P|}{s-|P|}}{\binom{n-1}{s}}}_{=C_\beta^{-1}}=1,
\quad\text{and}\quad
C^{(2)}_\beta = C_\beta\sum_{s=0}^{n-1}\frac{\beta(s)\binom{n-2-|P|}{s-|P|-1}}{\binom{n-1}{s}}.
\]
In the particular case where $\beta(s)$ is concentrated at $s=\bar s$, we have
\[
C^{(2)}_\beta = \frac{\binom{n-2-|P|}{\bar s-|P|-1}}{\binom{n-1-|P|}{\bar s-|P|}} = \frac{\bar s-|P|}{n-1-|P|}.
\]
\end{example}

The practical utility of Theorem~\ref{t:efficient_semivalue} relies on~$\Theta$ being of limited size and the terms $\bigl|\pc^s_u(\theta)\bigr|$ and $\bigl|\nc^s_u(\theta)\bigr|$ being efficiently computable, as in Example~\ref{e:example_int_edges2}. Both requirements are satisfied by the group centralities considered in this paper, guaranteeing the framework's applicability.

\section{Cardinality-Constrained Set-Function Maximization}
\label{s:theory_max}

The facility-building Example~\ref{e:hospitals} introduced in Section~\ref{s:prob_gtc} suggests game-theoretic centrality as a promising metric for identifying the best nodes to form a collaborative group, and the past-aware modification of GTC introduced in Section~\ref{s:def} accounts for already partially formed groups.

We now explore this approach further, extending the scenario of Example~\ref{e:hospitals} by introducing a PAGTC-based heuristic for the broad problem of maximizing a general set function over the nodes of a graph. Subsequently, we demonstrate the practical viability of our heuristic by applying it to two distinct, tractable classes of combinatorial optimization problems.

\subsection{PAGTC for set-function maximization}

Given a graph $G=(V,E)$, a set function $f:2^V\to\mathbb{R}$, and a \emph{budget} $0<r<|V|$, consider the optimization problem
\begin{equation}
\label{e:maxf}
\argmax_{S\subseteq V}\{f(S):|S|=r\}.
\end{equation}
A standard greedy strategy (Algorithm \ref{a:greedy}) is a conventional approach for finding an approximate solution to \eqref{e:maxf}. The greedy algorithm builds the solution $S_r$ by adding one node at a time. At the $k$-th step, it adds the node $u_k$ which maximizes the gain to the already-selected set of nodes $S_{k-1}$.
\begin{algorithm}
\caption{Greedy algorithm for Problem~\eqref{e:maxf}.}
\label{a:greedy}
\begin{algorithmic}
\State $S_0 \gets \varnothing$
\For {$k\in 1,\dots,r$}
    \State $u_k \gets\displaystyle\argmax_{u\notin S_{k-1}}f(u|S_{k-1})$
    \State $S_k \gets S_{k-1}\cup\{u_k\}$
\EndFor
\State\Return $S_r$
\end{algorithmic}
\end{algorithm}

In our approach, we build the solution by adding one node at a time, as in the greedy algorithm. However, at step $k$, instead of maximizing the local marginal gain, we propose maximizing the \emph{expected marginal contribution} of $u$, conditioned on the nodes in $S_{k-1}$ having already been selected. Specifically, we choose
\[
u_k\in\argmax_{u\notin S_{k-1}}\,\mathbb{E}\bigl[f(u|\bm{S})\mid S_{k-1}\subseteq \bm{S}\subseteq V\setminus\{u\}\bigr],
\]
where $\bm{S}$ is a random subset of nodes. It remains to determine the distribution from which to sample~$\bm{S}$. Since the final solution will have size $r$, each node $u_k$ will collaborate with $r-1$ other nodes. Therefore, a natural choice is to sample~$\bm{S}$ uniformly at random from the subsets of size $r-1$. In this case, the expected marginal contribution can be directly expressed in terms of past-aware game-theoretic centrality, since we have
\[
\mathbb{E}\bigl[f(u|\bm{S})\mid S_{k-1}\subseteq \bm{S}\subseteq V\setminus\{u\},\,|\bm S|=r-1 \bigr]
=\phi_{f}^{\delta_{r-1}}(u|S_{k-1}),
\]
where $\phi_{f}^{\delta_{r-1}}(u|S_{k-1})$ is the $\beta$-semivalue PAGTC (Definition~\ref{d:pagtc}) with the support of~$\beta(s)$ concentrated at $s=r-1$. The approach is summarized in Algorithm~\ref{a:semi}.
\begin{algorithm}
\caption{PAGTC approach for Problem~\eqref{e:maxf}.}\label{a:semi}
\begin{algorithmic}
\State $S_0 \gets \varnothing$
\For {$k=1,\dots,r$} 
    \State $u_k \gets\displaystyle\argmax_{u\notin S_{k-1}}\phi^{\delta_{r-1}}_f(u|S_{k-1})$
    \State $S_k \gets S_{k-1}\cup\{u_k\}$
\EndFor
\State\Return $S_r$
\end{algorithmic}
\end{algorithm}

At the first iteration, we have no information about which nodes $u$ will collaborate with, and $\phi_f ^{\delta_{r-1}}(u|S_0)$ coincides with $\phi_f^{\delta_{r-1}}(u)$, the standard semivalue GTC. As the algorithm progresses, the estimate of the~$r-1$ collaborators of~$u$ becomes increasingly accurate. Eventually, at the last step, $r-1$ nodes have been added to the solution set, and the final one is chosen deterministically, since $\phi_f ^{\delta_{r-1}}(u|S_{r-1})=f (u|S_{r-1})$, as noted in Example~\ref{e:beta_concentrated}.

\begin{example}
Consider the maximization problem~\eqref{e:maxf} with $f(S)=f_{\mathrm{den}}(S)$, introduced in Examples~\ref{e:example_int_edges} and~\ref{e:example_int_edges2}. This is known as the $r$-densest subgraph problem, which asks for a set $S$ of $r$ vertices that maximizes the number of edges with both ends in $S$. At step $k$ of the PAGTC algorithm, we add to the partial solution $S_{k-1}$ the node $u_k$ that maximizes the corresponding semivalue past-aware game-theoretic centrality, which we computed in Example~\ref{e:example_int_edges2}:
\[
\phi^{\delta_{r-1}}_{f_{\mathrm{den}}}(u|S_{k-1}) = \bigl|N(u)\cap S_{k-1}\bigr| + \frac{r-1-|S_{k-1}|}{n-1-|S_{k-1}|}\bigl|N(u)\setminus S_{k-1}\bigr|.
\]
While in this case the method results in a straightforward heuristic that essentially favors nodes with high degrees, the structure of the PAGTC expression is instructive, as it reflects the underlying principle of Algorithm~\ref{a:semi}. The term $|N(u)\cap S_{k-1}|=f_{\mathrm{den}}(u|S_{k-1})$ is purely greedy and measures the immediate increase in $f_{\mathrm{den}}$ when adding $u$ to $S_{k-1}$. The term $|N(u)\setminus S_{k-1}|$, on the other hand, accounts for the ``future'': selecting nodes with a high degree, even if they do not impact the current step, may be advantageous in later steps. The weight of this latter term is highest at the beginning, and then decreases until the last step, where $|S_{k-1}|=r-1$ and the algorithm performs a purely greedy step (see Example~\ref{e:beta_concentrated}).
\end{example}

\begin{remark}
\label{r:greedy_recover}
In Algorithm~\ref{a:semi}, if we instead choose the probability distribution $\beta = \delta_{k-1}$ at each step~$k$, we obtain exactly the greedy strategy of Algorithm~\ref{a:greedy} (see Example~\ref{e:beta_concentrated}).
\end{remark}

Before empirically evaluating the algorithm on two different classes of objective functions~$f$, we dedicate the following section to a theoretical analysis of its solution quality under certain structural assumptions.

\subsection{Optimality bounds in submodular case}
\label{s:optimality_bounds}

The standard greedy algorithm is known to perform well for a class of objective functions $f$ called \emph{submodular} functions.

\begin{definition}
We say that a set function $f:2^V\to\mathbb{R}$ is \emph{submodular} if $f(u|S)\ge f(u|T)$, for any two sets $S\subseteq T$ and any $u\notin T$. We say that $f$ is \emph{monotone} if $f(u|S)\ge0$, for any $u\in V$ and $S\subseteq V$.
\end{definition}

It has been shown in~\cite[Thm. 4.2]{nemhauser1978} that, if $f$ is non-negative, monotone, submodular, and $f(\varnothing)=0$, then the greedy solution $S$ of cardinality $r$ satisfies
\begin{equation}
\label{e:opt_bound_greedy}
f(S)\ge\left[1-\left(1-\dfrac{1}{r}\right)^{\!r\,}\right]f(S^*)\ge \left(1-\frac1e\right)f(S^*),
\end{equation}
where $S^*$ is an optimal solution and $e$ is Euler’s number. The approximation ratio of $1-1/e$ achieved by the greedy algorithm has been proven optimal for any algorithm restricted to a polynomial number of value queries~\cite{NemhauserOptimality}.

While this context is particularly favorable to the greedy algorithm, we establish a {weaker} theoretical bound for the PAGTC heuristic. Nevertheless, this analysis provides a rigorous performance guarantee and additionally yields computable \emph{a posteriori} bounds, which can be useful in practice to assess solution quality.

\begin{theorem}
\label{t:magic_bound}
If $f$ is non-negative, monotone, submodular and $f(\varnothing)=0$, then the solution $S_r$ of the PAGTC Algorithm~\ref{a:semi} to Problem~\eqref{e:maxf}, with optimal solution $S^*$, satisfies
\begin{equation}
\label{e:opt_bound_pagtc}
f(S_r)\ge \frac{r}{2r-1}f(S^*)\ge \dfrac{1}{2}f(S^*)
\end{equation}
\end{theorem}
\begin{proof}
The proof strategy is similar to~\cite[Thm. 4.2]{nemhauser1978}, although here we must account for expected marginal contributions. We begin by defining the random variables $\{\bm R_k\}_{k=1}^{r}$, where $\bm R_k$ is a subset of $r-k$ elements chosen uniformly at random from $V\setminus S_{k-1}$. For convenience, we also set $\bm T_k= \bm R_k\cup S_{k-1}$.

We recall \cite[Prop. 2.1 (v)]{nemhauser1978} that, for a submodular function $f$ and any $S$ and $T$, it holds:
\begin{equation}
\label{e:prop2.4_nemhauser1978}
f(S\cup T)\le f(T)+\sum_{u\in S\setminus T}f(u|T)
\end{equation}
and therefore, for $1\le k\le r$ we can write
\begin{equation}
\label{e:firstS*}
f(S^*)\le \mathbb E\bigl[f(S^*\cup \bm T_k)\bigr]
\le\mathbb E\bigl[f(\bm T_k)\bigr]+\sum_{u\in S^*}\mathbb E\bigl[f(u|\bm T_k)\bigr],
\end{equation}
where the first inequality follows from monotonicity. We can express $\mathbb E\bigl[f(u|\bm T_k)\bigr]$ in terms of PAGTC. Indeed, for $u\notin S_{k-1}$, we have
\begin{equation}
\label{e:bridge_E_phi}
\mathbb E\bigl[f(u|\bm T_k)\bigr] 
=\mathbb P[u\notin \bm R_k]\cdot\mathbb E\bigl[f(u|\bm T_k)\mid u\notin \bm R_k\bigr] = \alpha_k\cdot \phi^{\delta_{r-1}}_f(u|S_{k-1}),
\end{equation}
where we define
\[
\alpha_k=\frac{n-r+1}{n-k+1}.
\]
The node $u_k$, chosen by Algorithm~\ref{a:semi} to maximize $\phi^{\delta_{r-1}}_f(u|S_{k-1})$, then also maximizes $\mathbb E[f(u|\bm T_k)]$. Consequently, we find
\begin{equation}
\label{e:uk_vs_optimal}
 \sum_{u\in S^*} \mathbb E[f(u|\bm T_k)] \le r\,\mathbb E\bigl[f(u_k|\bm T_k)\bigr]=r\,\alpha_k\, p_k,
\end{equation}
where we set $p_k=\phi^{\delta_{r-1}}_f(u_k|S_{k-1})$.

We now evaluate the term $\mathbb E\bigl[f(\bm T_k)\bigr]$ of~\eqref{e:firstS*}, conditioning on whether $u_k\in \bm R_k$:
\begin{equation}
\label{e:E[tk]_conditioned}
\mathbb E\bigl[f(\bm T_k)\bigr]
=(1-\alpha_k)\,\mathbb E\bigl[f(\bm T_k)\mid u_k\in \bm R_k\bigr]+\alpha_k\, \mathbb E\bigl[f(\bm T_k)\mid u_k\notin \bm R_k\bigr]
\end{equation}
If $u_k\in \bm R_k$, then both $\bm T_k$ and $\bm T_{k+1}$ are uniform random completions of $S_k$ of cardinality $r-1$, and therefore $\mathbb E\bigl[f(\bm T_k)\mid u_k\in \bm R_k\bigr]=\mathbb E[f(\bm T_{k+1})]$. Conversely, when $u_k\notin \bm R_k$, by definition of $p_k$ we have
\[
p_k= \underbrace{\mathbb E\bigl[f(
\overbrace{
    \bm T_k\cup\{u_k\}
}^{=\bm R_k \cup S_k}
)\mid u_k\notin \bm R_k\bigr]}_{(*)}-\mathbb E\bigl[f(\bm T_k)\mid u_k\notin \bm R_k\bigr] .
\]
To evaluate $(*)$, note that for a fixed node $u\notin S_k$ we have
\[
\begin{split}
\phi^{\delta_{r-1}}_f(u|S_k) = \mathbb E\bigl[f(
\underbrace{
    \bm T_{k+1}\cup\{u\}
}_{\bm R_{k+1}\cup\{u\}\cup S_k}
)\mid u\notin \bm T_{k+1}\bigr]-\mathbb E\bigl[f(\bm T_{k+1})\mid u\notin \bm T_{k+1}\bigr].
\end{split}
\]
Notice that conditionally, $\bm T_{k+1} \cup \{u\}$ is a uniformly random $r$-superset of $S_k$ forced to contain $u$. Averaging this over all uniformly chosen nodes $u \notin S_k$ recovers the distribution of $\bm T_k\cup\{u_k\}$ in $(*)$. Consequently, averaging the entire identity over all $u\notin S_k$ yields
\[
q_k:=\dfrac{1}{n-k}\sum_{u\notin S_k}\phi^{\delta_{r-1}}_f(u|S_k)
= (*) - \mathbb E\bigl[f(\bm T_{k+1})\bigr],
\]
where the conditioning on $u\notin \bm T_{k+1}$ collapses when averaging $\mathbb E\bigl[f(\bm T_{k+1})\mid u\notin \bm T_{k+1}\bigr]$.
Substituting back into~\eqref{e:E[tk]_conditioned} gives
\[
\begin{split}
\mathbb E\bigl[f(\bm T_k)\bigr]
&=(1-\alpha_k)\,\mathbb E[f(\bm T_{k+1})]+\alpha_k\, \bigl(
\mathbb E\bigl[f(\bm T_k)\bigr]+q_k-p_k
\bigr)\\
& = \mathbb E[f(\bm T_{k+1})]+\alpha_k(q_k-p_k)
\end{split}
\]
and therefore, unrolling the recurrence, we obtain
\begin{equation}\label{e:EfTk_expression}
\mathbb E\bigl[f(\bm T_k)\bigr] = \underbrace{\mathbb E\bigl[f(\bm T_r)\bigr]}_{=f(S_{r-1})}+\sum_{i=k}^{r-1}\alpha_{i}(q_i-p_i).
\end{equation}
Note that $\alpha_iq_i\le \alpha_{i+1}p_{i+1}$, because $\alpha_i\le\alpha_{i+1}$ and the selection strategy of Algorithm~\ref{a:semi} ensures $\phi_f^{\delta_{r-1}}(u|S_i)\le \phi_f^{\delta_{r-1}}(u_{i+1}|S_i)$ for any $u\notin S_i$. Hence we obtain the following bound:
\begin{equation}
\label{e:bound_e[fT_k]}\mathbb E\bigl[f(\bm T_k)\bigr]\le
f(S_{r-1})+\alpha_rp_r-\alpha_kp_k=f(S_r)-\alpha_kp_k.
\end{equation}
where $\alpha_rp_r=f(u_r|S_{r-1})=f(S_r)-f(S_{r-1})$ follows from~\eqref{e:bridge_E_phi}.

Finally, substituting~\eqref{e:bound_e[fT_k]} and~\eqref{e:uk_vs_optimal} into~\eqref{e:firstS*}, we obtain
\[
f(S^*)\le f(S_r)+(r-1)\alpha_kp_k\le  f(S_r)+(r-1)p_k.\]
Since this holds at any step $k$, we can write $f(S^*)\le f(S_r)+(r-1)\hat p$, where $\hat p=\min_{k}p_k$. Furthermore, by submodularity we have
\[
f(S_r)=\sum_{k=1}^{r}f(u_k|S_{k-1}) \ge \sum_{k=1}^{r}\phi_f^{\delta_{r-1}}(u_k|S_{k-1})\ge r\hat p
\]
and therefore
\[
f(S^*)\le f(S_r)+\dfrac{r-1}r f(S_r) = \dfrac{2r-1}{r}f(S_r),
\]
which concludes the proof.
\end{proof}

\begin{remark}\label{r:a_posteriori_bounds}
Assuming that $|V|\ge 2r-1$, 
from~\eqref{e:firstS*} and~\eqref{e:bridge_E_phi}, we obtain the following data-dependent \emph{a posteriori} bound:
\[
f(S^*)\le\min_{1 \le k \le r} \left( \mathbb E\bigl[f(\bm T_k)\bigr]+\alpha_k\sum_{i=1}^r \phi_f^{\delta_{r-1}}\left(v_{i}^{(k)}\Big|S_{k-1}\right)\right),
\]
where $v_1^{(k)},\dots,v_r^{(k)}$ are the nodes in $V\setminus S_{k-1}$ yielding the $r$ largest values of the PAGTC metric $\phi_f^{\delta_{r-1}}(\,\cdot\,|S_{k-1})$. Note that the term $\mathbb E\bigl[f(\bm T_k)\bigr]$ can be evaluated using quantities already computed by the algorithm via~\eqref{e:EfTk_expression}.

This resembles the analogous \emph{a posteriori} bound for the greedy algorithm, which is a direct consequence of~\eqref{e:prop2.4_nemhauser1978}:
\[
f(S^*) \le \min_{1 \le k \le r} \left( f(S_{k-1}) + \sum_{i=1}^r f\left(v_i^{(k)}\Big|S_{k-1}\right) \right),
\]
where $v_1^{(k)},\dots,v_r^{(k)}$ are the nodes in $V\setminus S_{k-1}$ yielding the $r$ largest marginal contributions to $S_{k-1}$.

These bounds are useful for assessing the solution quality of either algorithm by comparing the obtained value $f(S_r)$ against an upper bound on the optimal value~$f(S^*)$. Depending on the problem instance, either bound could prove tighter in practice.
\end{remark}

\section{Two Application Examples}
\label{s:threee_examples}

As shown in Section~\ref{s:theory_max}, the PAGTC approach can be adopted to provide heuristics for maximizing a given objective function $f(S)$ over subsets of nodes $S \subseteq V$ of cardinality $r$. The practical feasibility of this approach is limited by the computational complexity of the PAGTC scores $\phi_f^{\delta_{r-1}}$, which must be computed for all nodes in the graph at every step of the algorithm.

In this section, we present two classes of problems, corresponding to distinct objective functions~$f$, which cover both the submodular and non-submodular cases. For each function, we  leverage the framework provided by Theorem~\ref{t:efficient_semivalue} to derive efficient closed-form expressions for the corresponding PAGTC scores.

\subsection{Facility location}
\label{s:example_rho}
We begin with problems of the form~\eqref{e:maxf}, with objective function
\[
f_\rho(S)=\begin{cases}
    \displaystyle\sum_{v \in V}\rho\bigl(d(S,v)\bigr) & \text{if $S\ne\varnothing$}\\[.5ex]
    0 & \text{if $S=\varnothing$},
\end{cases}
\]
parameterized by a function $\rho:\mathbb{N}\to\mathbb R$. Here, the importance of a group of nodes~$S$ is assessed based on how close every node in the graph is to~$S$, as specified by the function $\rho$.

\begin{remark}
\label{r:rho_nonincreasing}
If $\rho$ is nonnegative and nonincreasing, we have
\[
\rho\bigl(d(S,v)\bigr)=\rho\left(\min_{w\in S} d(w,v)\right)=\max_{w\in S}\rho\bigl(d(w,v)\bigr)
\]
and therefore, setting $c_{vw}=\rho\bigl(d(w,v)\bigr)\ge 0$, we can write $f_\rho$ in the form of the standard \emph{facility location} problem:
\[
f_\rho(S)=\sum_{v\in V}\max_{w\in S}c_{vw},
\]
which is known to be nonnegative, monotone and submodular~\cite{Krause_Golovin_2014}.
\end{remark}
The optimality performance bounds~\eqref{e:opt_bound_greedy} of the greedy algorithm for this class of problems were first studied in~\cite{NemhauserBank}, where the optimization of bank locations in cities is discussed. Several speed-ups over the standard greedy algorithm followed, such as the \emph{accelerated greedy algorithm}~\cite{AcceleratedGreedy} and the
CELF algorithm introduced for sensor placement in networks~\cite{CELF}.

Note that taking $\rho(x)=1/x$ for $x\ge1$ and $\rho(0)=0$, we obtain the \emph{group harmonic centrality} introduced in Example~\ref{e:hospitals}, which is \emph{not monotone}. The specific problem of maximizing group harmonic centrality, as well as other variants, is treated extensively in~\cite{angriman}.

In order to derive an efficient formula for computing the PAGTC scores for ${f_\rho}$ and apply our framework, we express $f_\rho(S)$ in the form of~\eqref{e:efficient_nu}, as done in~\cite{tarkowski2018}: we set $\Theta=V\times \mathbb N$ and $g(v,l)=\rho(l)$, and we define the relation $\sim$ such that $(v,l)\sim S$ whenever $l=d(S,v)$. This allows the application of Theorem~\ref{t:efficient_semivalue} and leads to the following result. Note that here we make no prior assumptions about the function~$\rho$.
\begin{theorem}
\label{t:formula_semivalue_rho}
Given a graph $G=(V,E)$ and a subset of nodes $P$, with $|P|<r$, for any $v\in V$ and any $l\in\mathbb{N}$, let $m^>(v,l)$ denote the number of nodes $w\in V$ such that $d_{vw}>l$. Then, for a node $u \notin P$, the $\beta$-semivalue PAGTC associated with $f_\rho$ (using $\beta=\delta_{r-1}$) is given by the following expression:
\begin{equation}
\label{e:formula_semivalue_rho}
\phi_{f_\rho}^{\delta_{r-1}}(u|P) = \sum_{v\notin P} \sum_{l=d_{uv}}^{d(P, v)-1} \Delta\rho(l)\frac{\binom{m^>(v,l)-|P|}{r-1-|P|}}{\binom{n-1-|P|}{r-1-|P|}}
\end{equation}
where $\Delta\rho(l)=\rho(l)-\rho(l+1)$.
\end{theorem}
\begin{remark}
The result holds when $P=\varnothing$, with the convention that $d(\varnothing, v)=\infty$, provided we set $\rho(l)=0$ for $l$ greater than the diameter of $G$.
\end{remark}
\begin{proof} Since the case $P=\varnothing$ has already been addressed in~\cite{tarkowski2018}, here we assume $P\ne\varnothing$ for simplicity.
Consider the positive contributions~\eqref{e:pc_and_nc} for a given node-distance pair $(v,l)$:
\[
\pc^s_u(v,l) = \{S\subseteq V\setminus\{u\}: |S|=s,\,P\subseteq S,\,d(v,S\cup\{u\})=l,\, d(v,S)>l\}.
\]
We note that $d_{uv}=l$ and $d(P,v)>l$ are necessary conditions for $\pc_u^s(v,l)$ to be non-empty. Under these assumptions, $S\in\pc^s_u(v,l)$ if and only if the $s-|P|$ elements of $S\setminus P$ are at a distance greater than $l$ from $v$; thus, we obtain
\def\mbin#1{\binom{#1-|P|}{s-|P|}}
\[
\bigl|\pc^s_u(v,l)\bigr|=\mbin{m^>(v,l)}.
\]

On the other hand, the set of negative contributions
\[
\nc^s_u(v,l) = \{S\subseteq V\setminus\{u\}: |S|=s,\,P\subseteq S,\,d(v,S\cup\{u\})<l,\, d(v,S)=l\}
\]
requires that $d_{uv}<l$ and $d(P,v)\ge l$. Under these assumptions, analogously to the positive contributions, we have
\[
\bigl|\nc^s_u(v,l)\bigr|=\begin{cases}
\mbin{m^\ge(v,l)}-\mbin{m^>(v,l)} & \text{if $d(P,v)>l$}\\[1.5ex]
\mbin{m^\ge(v,l)} & \text{if $d(P,v)=l$,}
\end{cases}
\]
where $m^\ge(v,l)$ denotes the number of nodes $w\in V$ such that $d_{vw}\ge l$. Note that when $d(P,v)>l$, computing the valid choices for $S\setminus P$ via this subtraction ensures that $S\setminus P$ contains at least one node $w$ with $d_{wv}=l$, and thus $d(v,S)=l$.

Recognizing that the positive contribution occurs only when $l = d_{uv}$, we can combine the expressions for $|\pc^s_u(v,l)|$ and $|\nc^s_u(v,l)|$ to obtain
\[
\begin{split}
&\sum_{l\in \mathbb N}\rho(l)\Bigl(\bigl|\pc^s_u(v,l)\bigr|-\bigl|\nc^s_u(v,l)\bigr|\Bigr)\\
&=\sum_{l=d_{uv}}^{d(P, v)-1}\!\!\rho(l)\mbin{m^>(v,l)}
-\sum_{l=d_{uv}+1}^{d(P, v)}\!\!\rho(l)\mbin{m^\ge(v,l)}\\
&= \sum_{l=d_{uv}}^{d(P, v)-1}\Delta\rho(l)\mbin{m^>(v,l)},
\end{split}
\]
where we used the relation $m^\ge(v,l)=m^>(v,l-1)$ and re-indexed the second sum. Then  Theorem~\ref{t:efficient_semivalue} gives the expression for the general semivalue:
\begin{equation}
\label{e:formula_general_semivalue_rho}
\phi_{f_\rho}^\beta(u|P) = C_\beta \sum_{v\notin P} \sum_{l=d_{uv}}^{d(P, v)-1} \Delta\rho(l)\sum_{s=0}^{n-1}\frac{\beta(s)}{\binom{n-1}{s}}\binom{m^>(v,l)-|P|}{s-|P|},
\end{equation}
where $C_\beta$ is given by~\eqref{e:c_beta}, from which~\eqref{e:formula_semivalue_rho} follows immediately.
\end{proof}
The corresponding PAGTC procedure for the facility problem is detailed in Algorithm~\ref{a:pagtc_facilities}, which has the same time complexity as the standard greedy algorithm, $\mathcal O(|V||E|+r|V|^2)$.
\begin{algorithm}
\caption{PAGTC approach for facility location.}
\label{a:pagtc_facilities}
\begin{algorithmic}
\For { $v \in V$ }
    \Precompute $d_{vw}$ for each $w\in V$ and $m^>(v,l)$ for each $l\le\diam(G)$ via BFS
\EndFor
\State\hh{12pt} $S_0 \gets \varnothing$
\For{\hh{12pt} $k = 1,\dots,r$}
    \For {$m= 0,\dots,n$}
        \State\hh{12pt} $F_k(m)\gets\binom{m-|S_{k-1}|}{r-1-|S_{k-1}|}/\binom{n-1-|S_{k-1}|}{r-1-|S_{k-1}|}$
    \EndFor\vspace{1ex}
    \ForAll {$v\in V\setminus S_{k-1}$ and $x\le\diam(G)$}
        \State\hh{12pt} $G_k(v,x)\gets\sum_{l=x}^{d(S_{k-1}, v)-1} \Delta\rho(l)F_k\bigl(m^>(v,l)\bigr)$
    \EndFor\vspace{1ex}
    \For{ $u\in V\setminus  S_{k-1}$}
        \State\hh{12pt} $\phi^{\delta_{r-1}}_{f_\rho}(u|S_{k-1})\gets\sum_{v\notin S_{k-1}}G_k(v,d_{uv})$
    \EndFor\vspace{1ex}
    \State $u_k \gets\displaystyle\argmax_{u\notin S_{k-1}}\phi^{\delta_{r-1}}_{f_\rho}(u|S_{k-1})$
    \State\hh{11pt} $S_k \gets S_{k-1}\cup\{u_k\}$
\EndFor
\State\hh{11pt} \Return $S_r$
\end{algorithmic}
\end{algorithm}

The bounds for the greedy and PAGTC algorithms described in Section~\ref{s:optimality_bounds} hold, provided that $\rho$ is nonnegative and nonincreasing. In Section~\ref{s:distance_experiments}, we evaluate both the quality of the solution and the \emph{a posteriori} bounds.

\subsection{Influence maximization in complex contagion}
\label{s:example_K}
As a second example, we address the problem of \emph{influence maximization} under the dynamics of {complex contagion}. Specifically, in $K$-\emph{complex contagion}, reinforcement from $K$ neighbors is required for the transmission of an infection or the spread of information. We adopt the latter interpretation, as our goal is to study strategies that \emph{maximize} the overall propagation of the cascade.

Under this model, each node is classified as either \emph{active} or \emph{inactive}. Once active, a node cannot revert to an inactive state. Starting from an initial set of active nodes, the dynamics proceed in rounds: at each round, any inactive node with at least~$K$ active neighbors becomes active. The cascade may reach a fixed point before the entire graph is activated.

The following group centrality can be naturally associated with the dynamics~\cite[Sec. 4.2]{CalculationShapley}:
\[
f_K(S) = |S|+\Bigl|\bigl\{u\in V\setminus S: |N(u)\cap S|\ge K\bigr\}\Bigr|.
\]
When $S$ is the set of currently active nodes, $f_K(S)$ represents the total number of nodes that will be active in the \emph{next round} of the $K$-complex contagion process.  This group centrality was originally introduced in~\cite{CalculationShapley} within the context of game-theoretic centrality, where a closed-form expression for the corresponding Shapley GTC was derived. Because $f_K$ can be interpreted as ``looking one step ahead'' in the complex contagion dynamics, we refer to it as the \emph{one-round influence} of $S$, and we term the corresponding maximization problem~\eqref{e:maxf} \emph{one-round influence maximization}.

While $f_K(S)$ captures the immediate momentum of the spread, a harder and more interesting problem is \emph{full influence maximization}, which evaluates the full cascade. Let $f_K^*(S)$ denote the total number of active nodes once the process reaches its fixed point, given the initial seed set $S$. The goal of full influence maximization is to identify the seed nodes to be activated initially so that the final coverage, $f_K^*(S)$, is maximized.

Most literature on influence maximization focuses on other diffusion models, such as the independent cascade and linear threshold models~\cite{kempe2003}. These models are characterized by a \emph{submodular} influence function, which ensures that a greedy strategy, along with various algorithmic improvements~\cite{IMinNearlyOptimalTime, Chen2010ScalableIM, IMGreedyImprovement}, yields an approximation that satisfies the optimality bound~\eqref{e:opt_bound_greedy}.

\begin{figure}
\centering
\tikzgraphsettings
\begin{tikzpicture}[ellipse around/.style={empty, ellipse, draw, dashed}]
    \node (1) [label={[label distance=-2]above:$w$}] {};
    \node (2) [right of=1, label={[label distance=-2.5]above:$v$}] {};
    \node (3) [below of=2, label={[label distance=-2]left:$u$}] {};
    \node (4) [right of=3, label={[label distance=-2]right:$z$}] {};
    \draw (1) -- (4);
    \draw (2) -- (4);
    \draw (3) -- (4);
    \node[ellipse around, inner sep=4pt, fit=(2), label={[label distance=-2]150:$S$}] (S) {};
    \node[ellipse around, inner sep=10pt, fit=(1)(2), label=above:{$T$}] (T) {};
\end{tikzpicture}
\caption{Submodularity violation for $f_K$ and $f_K^*$ ($K=3$). In both cases, the marginal contribution of node $u$ is strictly greater for group $T$ than for group $S$, even though $S \subset T$.}
\label{f:not_submodular}
\end{figure}

As for complex contagion, game-theoretic approaches to influence maximization have been explored for the submodular case $K=1$~\cite{SuriNarahari, DiscountedShapley}. For $K>1$, however, the complex contagion process is \emph{non-submodular} under both the \emph{one-round} and \emph{full} influence measures (Figure~\ref{f:not_submodular}). This makes the problem substantially more challenging: existing approaches rely on dynamic programming~\cite{DontBeGreedy}, reinforcement learning~\cite{chen2023}, or analytical solutions restricted to specific random graph topologies~\cite{schoenebeck2022, ebrahimi2017, ebrahimi2015, ghasemiesfeh2018}. Because these methods often struggle with scalability or require strict structural assumptions, developing efficient heuristics for general, large-scale networks remains an active challenge.

\subsubsection{One-round influence maximization}
For our approach, we first focus on the one-round influence $f_K$ and derive an efficient formula for the corresponding PAGTC scores.

\begin{theorem}
\label{t:shapley_influence}
Given a graph $G=(V,E)$ and a subset of nodes $P$ with $|P|<r$, for any $v\in V$ let  $s_v$ denote the number of neighbors of $v$ outside $P$ and define
\[
r_v=K-1-(\deg(v)-s_v)=K-1-|N(v)\cap P|.
\]
Then, for a node $u \notin P$, the $\delta_{r-1}$-semivalue PAGTC associated with $f_K$ is given by the following expression:
\begin{equation}
\label{e:pagtc_influence_delta_formula}
\phi^{\delta_{r-1}}_{f_K}(u|P)=\hspace{-1ex}\sum_{j=0}^{\min\{s_u,r_u\}}\hspace{-1ex}C(s_u,j)
+\hspace{-1ex}\sum_{\substack{v\notin P,\,r_v\ge0\\v\in N(u)\\ \deg(v)\ge K}}\hspace{-.5ex}C(s_v,r_v)\left(1-\frac{r_v}{s_v}\right),
\end{equation}
where we define
\[
C(x,y)
=\dfrac{\binom{r-1-|P|}{y}\binom{n-r}{x-y}}{\binom{n-1-|P|}{x}}.
\]
\end{theorem}
\begin{proof}
We express $f_K$ in the form of~\eqref{e:efficient_nu} by setting $\Theta = V$, $g\equiv 1$, and by defining the relation $\sim$ such that $v\sim S$ if and only if $v\in S$ or there are at least $K$ neighbors of $v$ in $S$. Then Theorem~\ref{t:efficient_semivalue} and~\eqref{e:c_beta} give
\begin{equation}
\label{e:efficient_semivalue_proof_influence}
\phi^{\delta_{r-1}}_{f_K} (u|P)=
\dfrac{C_\beta}{\binom{n-1}{r-1}}
\sum_{v\in V}
\bigl|\pc^{r-1}_u(v)\bigr|
=\dfrac{1}{\binom{n-1-|P|}{r-1-|P|}}
\sum_{v\in V}
\bigl|\pc^{r-1}_u(v)\bigr|,
\end{equation}
as, in this context, the sets of negative contributions are always empty.

As for positive contributions $\pc_u^{r-1}(v)$, let $S$ be a group of $r-1$ nodes such that $P\subseteq S\subseteq V\setminus\{u\}$: given a node $v\ne u$, the group $S$ belongs to $\pc^{r-1}_u(v)$ if and only if $v$ is not in $S$, $u$ is a neighbor of $v$, and $S$ contains exactly $K-1$ neighbors of~$v$. Therefore, provided that $v\in N(u)$, $\deg(v)\ge K$, and $|N(v)\cap P|\le K-1$, the number  of choices for $S$ are
\begin{equation}
\label{e:pcv}
\left|\pc^{r-1}_u(v)\right|=
\overbrace{\underbrace{\binom{s_v-1}{r_v}}_{=\left(1-\frac{r_v}{s_v}\right)\binom{s_v}{r_v}}}^{\substack{\text{choices for}\\\text{the neighbors}\\ \text{of $v$ in $S\setminus P$}}}
\,\,\overbrace{\binom{n-1-|P|-s_v}{r-1-|P|-r_v}}^{\substack{\text{choices for the}\\\text{\emph{non-neighbors}}\\ \text{of $v$ in $S\setminus P$}}}.\\
\end{equation}

In the case where $v=u$, we have that $S$ belongs to $\pc^{r-1}_u(u)$ if and only if there are \emph{strictly less} than~$K$ neighbors of $u$ in $S$, leading to
\begin{equation}
\label{e:pcu}
\left|\pc^{r-1}_u(u)\right|=
\sum_{j=0}^{m}
\binom{s_u}{j}
\binom{n-1-|P|-s_u}{r-1-|P|-j},
\end{equation}
where $j$ represents the number of neighbors of $u$ in $S\setminus P$ and $m=\min\{s_u,r_u\}$.
Note that the case where $|N(u)\cap P|\ge K$ is handled by~\eqref{e:pcu}: in this scenario, $r_u<0$, and therefore $\left|\pc^{r-1}_u(u)\right|=0$.

The conclusion follows by substituting~\eqref{e:pcv} and~\eqref{e:pcu} into~\eqref{e:efficient_semivalue_proof_influence}, and then applying the combinatorial identity
\[
\frac{\binom{n-1-|P|-x}{r-1-|P|-y}\binom{x}{y}}{\binom{n-1-|P|}{r-1-|P|}}
=\dfrac{\binom{r-1-|P|}{y}\binom{n-r}{x-y}}{\binom{n-1-|P|}{x}},
\]
which can be verified by expanding the factorials.
\end{proof}

This yields Algorithm~\ref{a:pagtc_influence}, which runs in $\mathcal O(r|E|)$ time, matching the complexity of the standard greedy algorithm in the one-round influence maximization setting.

\begin{algorithm}
\caption{PAGTC for one-round influence maximization in complex contagion.}
\label{a:pagtc_influence}
\begin{algorithmic}
\State $S_0 \gets \varnothing$
\For{$k = 1,\dots,r$}
    \For {$v\in V\setminus S_{k-1}$ }
        \Precompute $s_v$, $r_v$ and $C(s_v,r_v)$
        \For  {$j=0,\dots,\min\{s_v,r_v\}\le \deg( v)$ }
            \Precompute $C(s_v, j)$
        \EndFor
    \EndFor
    \For { $u\in V\setminus S_{k-1}$ }
        \Compute $\phi^{\delta_{r-1}}_{f_K}(u|S_{k-1})$ using~\eqref{e:pagtc_influence_delta_formula}
    \EndFor
    \State \hh{11pt}$u_k \gets\displaystyle\argmax_{u\notin S_{k-1}}\phi^{\delta_{r-1}}_{f_K}(u|S_{k-1})$
    \State $S_k \gets S_{k-1}\cup\{u_k\}$
\EndFor
\State \Return $S_r$
\end{algorithmic}
\end{algorithm}

\subsubsection{Full-influence maximization}
\label{s:full-influence-maximization}

\enlargethispage{1.5ex}
The more challenging \emph{full-influence maximization} problem requires maximizing the final cascade size~$f_K^*$. The greedy approach here is more computationally expensive: as a single evaluation of $f_K^*$ requires simulating the full contagion cascade, the total time complexity rises to $\mathcal{O}(r|V||E|)$.

Applying the PAGTC approach directly to $f_K^*$ is not covered by the exact-computation framework of Section~\ref{s:computation_pagtc}, and we do not have an efficiently computable expression for the corresponding PAGTC scores. To circumvent this, we propose to leverage the seeds obtained from the one-round PAGTC algorithm as a candidate solution for the full cascade. Optimizing the first round of the evolution aims to prevent early activation bottlenecks. We empirically evaluate this approach in Section~\ref{s:experiments}, benchmarking our one-round PAGTC solutions against the more expensive full-influence greedy baseline.

\goodbreak 
\section{Numerical Experiments}
\label{s:experiments}

In this section, we evaluate numerically the PAGTC strategy discussed in Section~\ref{s:theory_max}, applying it to the problems outlined in Section~\ref{s:threee_examples}. The tests were conducted on an M1 Pro CPU with 16 GB of RAM, using Python 3.12.13 and NetworkX 3.4.2 \cite{networkx}.

\subsection{Facility location}
\label{s:distance_experiments}

We begin with the facility location problem introduced in Section~\ref{s:example_rho}, with objective function $f_\rho$. In our simulations, we consider the special case of \emph{maximum coverage problem}, where $\rho$ is defined as
\[
\rho(l)=\begin{cases}
1 & \text{if $l\le R$}\\
0 & \text{otherwise,}
\end{cases}
\]
for a given radius $R$, so that $f_\rho(S)$ counts the number of nodes at a distance of at most $R$ from $S$. Since $\rho$ is  nonnegative and nonincreasing, the resulting problem is submodular (Remark~\ref{r:rho_nonincreasing}).

\begin{figure}
\begin{subfigure}{.28\textwidth}
\centering
\includegraphics[height=3.5cm]{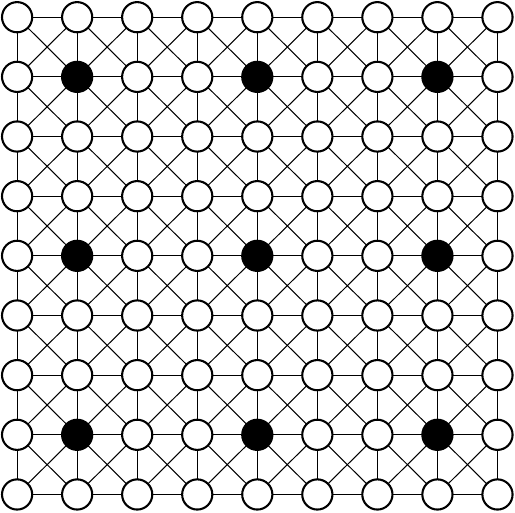}
\caption{\vphantom{$\phi_{f_\rho}^{\delta_{r-1}}(u|\varnothing)$}Optimal solution}
\label{f:king_opt}
\end{subfigure}\hfill\hfill
\begin{subfigure}{.33\textwidth}
\centering
\includegraphics[height=3.5cm]{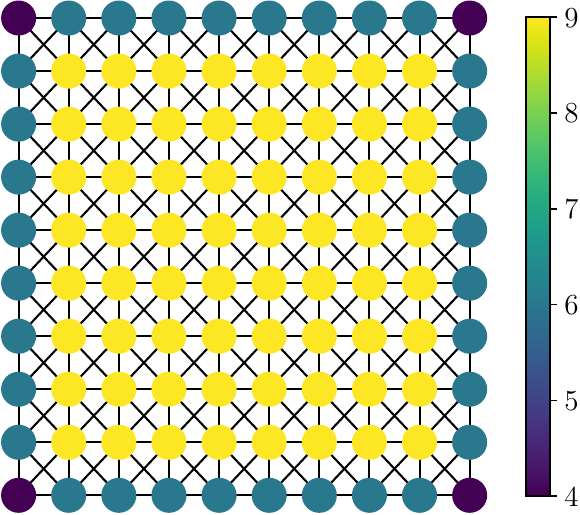}
\caption{\vphantom{$\phi_{f_\rho}^{\delta_{r-1}}(u|\varnothing)$}$f_\rho(u|\varnothing)$}
\label{f:king_greedy}
\end{subfigure}\hfill
\begin{subfigure}{.34\textwidth}
\centering
\includegraphics[height=3.5cm]{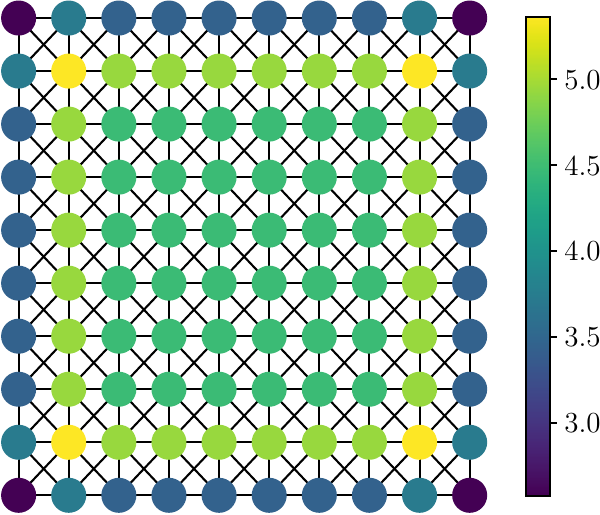}
\caption{$\phi_{f_\rho}^{\delta_{r-1}}(u|\varnothing)$}
\label{f:king_pagtc}
\end{subfigure}
\caption{Maximum coverage problem with radius $R=1$ and budget $r=9$.}
\label{fig:placeholder}
\end{figure}

\begin{table}
\centering
\caption{Coverage ratio (in \%) achieved by the greedy and PAGTC algorithms on King's graphs. The best result for each instance is highlighted in bold.}\vspace{-1ex}
\label{t:king}
$\begin{array}{*{10}c}
\toprule
R \text{ (radius)} & \multicolumn{3}{c}{2} & \multicolumn{3}{c}{3} & \multicolumn{3}{c}{4}\\
\cmidrule(lr){1-1}
\cmidrule(lr){2-4}
\cmidrule(lr){5-7}
\cmidrule(lr){8-10}
r \text{ (budget)} & 4 & 9 & 16  & 4 & 9 & 16  & 4 & 9 & 16  \\
 \midrule
 |V|=r(2R+1)^2 &100 & 225 & 400 & 196 & 441 & 784 & 324 & 729 & 1296 \\
 \midrule
\text{Greedy $(\%$)} & 95.0 & 84.9 & 87.0 & 93.4 & 84.4 & 87.8 & 86.1 & 82.7 & 86.2\\
\text{PAGTC ($\%$)} & \bm{100} & \bm{100} & \bm{91.7} & \bm{100} & \bm{100} & \bm{89.8} & \bm{100} & \bm{100} & \bm{88.9} \\
\bottomrule
\end{array}$
\end{table}

We first focus on instances where the optimal solution is straightforward to compute. Let $G_m$ be an $m\times m$ King's graph, i.e., a grid graph with additional edges between all diagonally adjacent nodes (see Figure~\ref{f:king_opt}). For a given coverage radius $R$, a single node can cover up to $(2R+1)^2$ nodes. Consequently, the entire graph can be perfectly covered by $r$ nodes whenever $|V|=r(2R+1)^2$, as exemplified in Figure~\ref{f:king_opt}. The symmetric topology of the graph provides a clear setting to compare the behavior of the greedy and PAGTC algorithms. In the first step, the greedy approach selects a node $u$ that maximizes the immediate marginal gain $f_\rho(u\vert{}\varnothing)=f_\rho\bigl(\{u\}\bigr)$. As Figure~\ref{f:king_greedy} illustrates, this choice is far from unique, since the vast majority of central nodes yield the exact same maximum gain. In contrast, the PAGTC algorithm selects its first node by maximizing the game-theoretic centrality $\phi_{f_\rho}^{\delta_{r-1}}(u\vert{}\varnothing)=\phi_{f_\rho}^{\delta_{r-1}}(u)$. This metric offers a richer structural evaluation that naturally breaks the initial symmetry, assigning the highest scores to four nodes belonging to the optimal solution (Figure~\ref{f:king_pagtc}). This pattern of behavior continues in subsequent steps. Finally, Table~\ref{t:king} compares the coverage ratios achieved by both algorithms across different values of $r$ and $R$, with the total number of nodes~$|V|$ chosen such that a perfect coverage is theoretically possible. To account for the arbitrary tie-breaking inherent in the greedy algorithm, the results are averaged over $10$ random reorderings of the graph's nodes.

While it is possible to construct several other instances where the PAGTC algorithm outperforms the greedy approach, such examples generally rely on highly symmetric topologies or rigidly constrained optimal configurations. In the majority of real-world graphs, the submodular nature of the objective function makes the standard greedy algorithm notoriously effective; as a result, both methods tend to produce very similar solutions. 

\begin{figure}
\begin{subfigure}{.43\textwidth}
\centering
    \includegraphics[width=\linewidth]{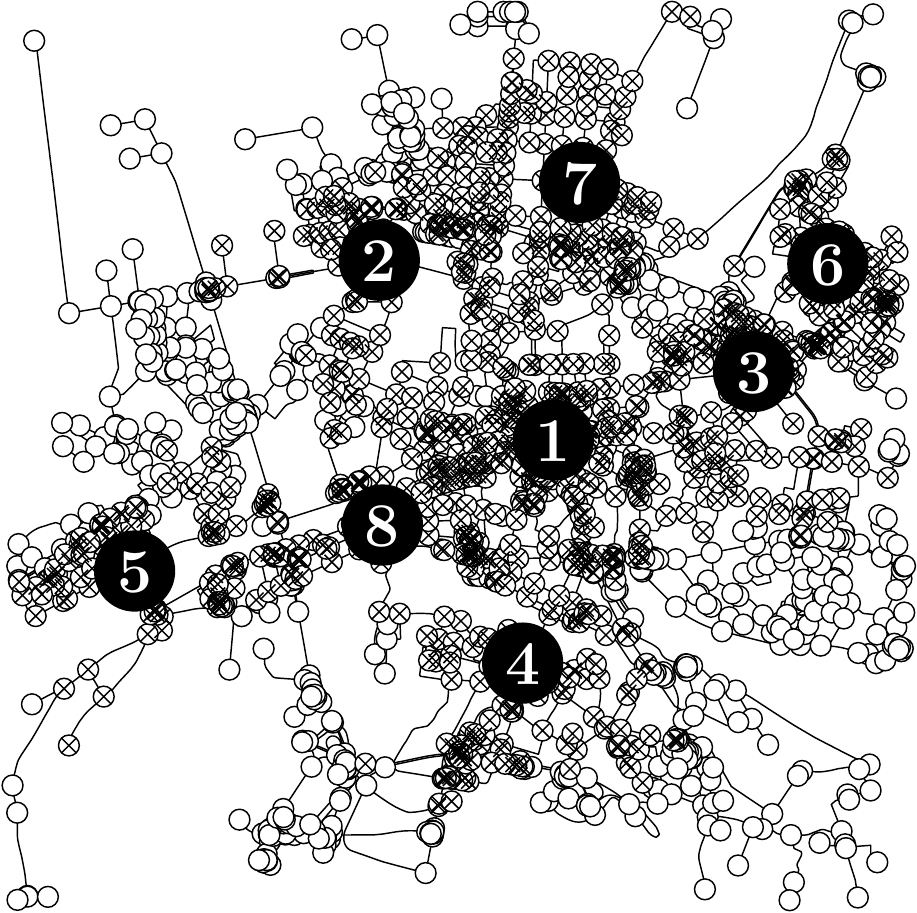}
    \caption{Identical solution selected by both the greedy ($S_G$) and PAGTC ($S_P$) algorithms for  $r = 8$. The numbers denote the selection sequence and the crossed nodes represent the covered nodes.}
    \label{f:map-bounds.map}
\end{subfigure}\hfill
\begin{subfigure}{.54\textwidth}
    \raggedleft
    \begin{tikzpicture}
    \begin{axis}[
        width=6cm,
        height=5.5cm,
        xtick distance=1,
        xticklabel style = {font=\footnotesize},
        yticklabel style = {font=\footnotesize},
        xlabel={$r$},
        legend style={at={(.9, 0.1)}, anchor=south west, font=\footnotesize},
    ]
    \addplot[color=mplblue, thick, mark=asterisk, mark options={scale=1.5}] table {figures/pisa_greedy_solution.txt};
    \addlegendentry{$f_\rho(S_{\mathrm{G}})$}
    \addplot[color=mplorange, thick, mark=square*, mark options={scale=0.6}] table {figures/pisa_pagtc_solution.txt};
    \addlegendentry{$f_\rho(S_{\mathrm{P}})$}
    \addplot[color=mplgreen, thick, mark=asterisk, mark options={scale=1.2, solid}, densely dotted] table {figures/pisa_greedy_bound.txt};
    \addlegendentry{$B_{\mathrm{G}}$}
    \addplot[color=mplred, thick, mark=square*, mark options={scale=0.6, solid}, densely dotted] table {figures/pisa_pagtc_bound.txt};
    \addlegendentry{$B_{\mathrm{P}}$}
    \addplot[color=mplpurple, thick, dashed] table {figures/pisa_lenG.txt};
    \addlegendentry{$|V|$}
    \end{axis}
    \end{tikzpicture}
    \caption{Comparison of the coverage values $f_\rho$ and the a posteriori upper bounds ($B_{\text{G}}$ for greedy, $B_{\text{P}}$ for PAGTC) as a function of the budget $r$. The dashed line marks the total number of nodes $|V|$. The graphs of $f_\rho(S_{\text{G}})$ and $f_\rho(S_{\text{P}})$ almost overlap.}
    \label{f:map-bounds.bounds}
\end{subfigure}
\caption{Coverage problem with radius $R = 9$ on the road network of Pisa, with edges of unit length.}
\label{f:map-bounds}
\end{figure}

Figure~\ref{f:map-bounds} illustrates this behavior for a coverage problem ($R=9$) on a real-world network. As the budget $r$ varies, the coverage values $f_\rho(S_G)$ and $f_\rho(S_P)$ overlap almost perfectly (Figure~\ref{f:map-bounds.bounds}). In some cases, both methods yield the exact same solution set, as exemplified for $r=8$ in Figure~\ref{f:map-bounds.map}.
However, even when the output sets are identical, the PAGTC framework may offer an analytical advantage: by exploiting the a posteriori bounds established in Remark~\ref{r:a_posteriori_bounds}, PAGTC can yield tighter upper bounds on the optimal solution compared to the standard greedy bounds. This phenomenon is visible in Figure~\ref{f:map-bounds.bounds}, where the PAGTC bound ($B_{\text{P}}$) remains tighter than the greedy bound ($B_{\text{G}}$) for higher values of $r$.

\subsection{One-round and full influence maximization}

We now shift our focus to the non-submodular context of influence maximization in complex contagion, evaluating both the one-round and full diffusion processes outlined in Section~\ref{s:example_K}. In our experiments, we consider the following graphs:
\begin{itemize}
    \item \texttt{small-world} denotes a set of $100$ undirected \emph{navigable small-world} networks~\cite{KleinbergSmallWorld}, each containing $25$ nodes and an average of about $52$ edges, generated using NetworkX’s \texttt{\lstinline{navigable_small_world}} function~\cite{networkx}.
    \item \texttt{flor-families} (15 nodes, 20 edges) is a well-known datasets available via NetworkX~\cite{networkx} representing marriage ties among Renaissance Florentine families.
    \item \texttt{retweets} (96 nodes, 117 edges) and \texttt{fb-messages} (1266 nodes, 6451 edges) represent retweets among Twitter users for selected hashtags and messages among a Facebook community of students at the University of California, Irvine~\cite{NetworkDataRepository}.
\end{itemize}

In Table~\ref{t:table_experiments}, we compare the greedy and PAGTC-based approaches for both one-round and full influence maximization. As detailed in Section~\ref{s:full-influence-maximization}, the full-influence evaluation benchmarks the computationally expensive greedy baseline (applied directly to the full cascade $f_K^*$) against the candidate seeds generated by the one-round PAGTC algorithm. The table reports the achieved influence as a percentage of the total number of nodes. The results are averaged over $10$ random reorderings of the nodes of the graph.

\def\.#1.#2{\textbf{#1.#2}}
\begin{table}
\caption{Comparison between greedy and PAGTC approaches for one-round and full influence maximization in $K$-complex contagion, for different values of $K$ and budget of~$r=2K$. We report the one-round and full influence of the solution, normalized (in \%). The optimal value (opt.), found by exhaustive enumeration, is reported for smaller graphs, and the best value, excluding the optimal one, is highlighted.}\par\vspace{-1ex}
\label{t:table_experiments}
\begin{tabular}{lc@{\quad\;}*3cc*3c}
\toprule
\multirow{3}*{Graph $G$} & \multirow{3}*{$K$} & \multicolumn{3}{c}{One-round influence} && \multicolumn{3}{c}{Full influence}\\\cline{3-5}\cline{7-9}\\[-2ex]
&& greedy & PAGTC & opt. && greedy & PAGTC & opt. \\
 && \small $\mathcal O(r|E|)$ & \small $\mathcal O(r|E|)$ &&& \small $\mathcal O(r|V||E|)$ & \small $\mathcal O(r|E|)$\\
\midrule
\multirow{3}*{\begin{tabular}{@{}l@{}}\texttt{small-world}\\(average of 100)\end{tabular}}
		&2& 41.8 & \.45.1 & 47.1 && \.100.0 &  98.1 & 100\\
		&3& 38.2 & \.45.1 & 48.0 && \.55.8 & 51.3 & 75.6\\
		&4& 37.8 & \.47.8 & 49.5 && 39.4 & \.49.2 & 56.6\\
\midrule
\multirow{3}*{\texttt{flor-families}}
		&2& \.56.7 & 60.0 & 66.7 && \.82.7 & 66.7 & 86.7\\
		&3& 52.0 & \.60.0 & 60.0 && 54.7 & \.66.7 & 73.3\\
		&4& 55.3 & \.73.3 & 73.3 && 55.3 & \.73.3 & 73.3\\
\midrule
\multirow{3}*{\texttt{retweets}}
		&2& 8.3 & \.12.5 & 13.5 && \.25.2 & 21.1 & 26.0\\
		&3& 6.7 & \.9.4 & 11.5 && 6.9 & \.10.2 & 15.6\\ 
		&4& 8.3 & \.11.5 & && 8.3 & \.11.5 & \\
\midrule
\multirow{4}*{\texttt{fb-messages}}
		&2& 3.9 & \.6.5 & && \.80.8 & 80.4 & \\
		&3& 0.9 & \.4.6 & && 41.4 & \.68.1 & \\ 
		&4& 0.6 & \.3.2 & && 0.6 & \.58.1 & \\
		&5& 0.8 & \.2.4 & && 0.8 & \.51.0 & \\
\bottomrule
\end{tabular}
\end{table}

\begin{figure}
\tikzgraphsettings
\begin{subfigure}{.33\textwidth}
\centering
\begin{tikzpicture}[scale=.85]
\node[fill=black] (0) at (0,0) {1};
\node[fill=black] (1) at (0,1) {2};
\node[fill=black] (2) at (1,0) {3};
\node[fill=black] (3) at (0,2) {5};
\node[fill=white,draw=none, label=center:\scriptsize$\times$, draw] (4) at (1,1) {};
\node[fill=black] (5) at (0,3) {6};
\node[fill=black] (6) at (1,2) {7};
\node[fill=white,draw] (7) at (0,4) {};
\node[fill=white,draw] (8) at (1,3) {};
\node[fill=white,draw] (9) at (1,4) {};
\node[fill=white,draw=none, label=center:\scriptsize$\times$, draw] (10) at (2,0) {};
\node[fill=black] (11) at (2,1) {4};
\node[fill=white,draw] (12) at (2,2) {};
\node[fill=white,draw] (13) at (2,3) {};
\node[fill=white,draw] (14) at (2,4) {};
\node[fill=white,draw] (15) at (3,0) {};
\node[fill=white,draw] (16) at (3,1) {};
\node[fill=white,draw] (17) at (3,2) {};
\node[fill=white,draw] (18) at (3,3) {};
\node[fill=white,draw] (19) at (3,4) {};
\node[fill=white,draw] (20) at (4,0) {};
\node[fill=white,draw] (21) at (4,1) {};
\node[fill=white,draw] (22) at (4,2) {};
\node[fill=white,draw] (23) at (4,3) {};
\node[fill=white,draw] (24) at (4,4) {};



\draw (0)  to (1);
\draw (0)  to (2);
\draw (1)  to (0);
\draw (1)  to (3);
\draw (1)  to (4);
\draw (2)  to (0);
\draw (2)  to (4);
\draw (2)  to (10);
\draw (2)  to (11);
\draw (3)  to (1);
\draw (3)  to (5);
\draw (3)  to (6);
\draw (4)  to (1);
\draw (4)  to (2);
\draw (4)  to (6);
\draw (4)  to (11);
\draw (4)  to (12);
\draw (5)  to (3);
\draw (5)  to (7);
\draw (5)  to (8);
\draw (5)  to (4);
\draw (6)  to (3);
\draw (6)  to (4);
\draw (6)  to (8);
\draw (6)  to (12);
\draw (7)  to (5);
\draw (7)  to (9);
\draw (8)  to (5);
\draw (8)  to (6);
\draw (8)  to (9);
\draw (8)  to (13);
\draw (9)  to (7);
\draw (9)  to (8);
\draw (9)  to (14);
\draw (9)  to[bend right=15] (2);
\draw (10)  to (2);
\draw (10)  to (11);
\draw (10)  to (15);
\draw (10)  to (1);
\draw (11)  to (4);
\draw (11)  to (10);
\draw (11)  to (12);
\draw (11)  to (16);
\draw (12)  to (6);
\draw (12)  to (11);
\draw (12)  to (13);
\draw (12)  to (17);
\draw (13)  to (8);
\draw (13)  to (12);
\draw (13)  to (14);
\draw (13)  to (18);
\draw (13)  to (2);
\draw (14)  to (9);
\draw (14)  to (13);
\draw (14)  to (19);
\draw (15)  to (10);
\draw (15)  to (16);
\draw (15)  to (20);
\draw (15)  to (11);
\draw (16)  to (11);
\draw (16)  to (15);
\draw (16)  to (17);
\draw (16)  to (21);
\draw (16)  to (12);
\draw (17)  to (12);
\draw (17)  to (16);
\draw (17)  to (18);
\draw (17)  to (22);
\draw (17)  to[bend right=15] (2);
\draw (18)  to (13);
\draw (18)  to (17);
\draw (18)  to (19);
\draw (18)  to (23);
\draw (18)  to (24);
\draw (19)  to (14);
\draw (19)  to (18);
\draw (19)  to (24);
\draw (19)  to (22);
\draw (20)  to (15);
\draw (20)  to (21);
\draw (21)  to (16);
\draw (21)  to (20);
\draw (21)  to (22);
\draw (22)  to (17);
\draw (22)  to (21);
\draw (22)  to (23);
\draw (22)  to (2);
\draw (23)  to (18);
\draw (23)  to (22);
\draw (23)  to (24);
\draw (23)  to (19);
\draw (24)  to (19);
\draw (24)  to (23);
\draw (24)  to[bend left=15] (20);
\end{tikzpicture}
\caption{Greedy: $f_K(S_{\text{G}})=9.$}
\label{f:max_greedy}
\end{subfigure}\qquad
\begin{subfigure}{.33\textwidth}
\centering
\begin{tikzpicture}[scale=.85]
\node[fill=white,draw] (0) at (0,0) {};
\node[fill=white,draw] (1) at (0,1) {};
\node[fill=white,draw=none, label=center:\scriptsize$\times$, draw] (2) at (1,0) {};
\node[fill=white,draw] (3) at (0,2) {};
\node[fill=white,draw] (4) at (1,1) {};
\node[fill=black] (5) at (0,3) {7};
\node[fill=white,draw] (6) at (1,2) {};
\node[fill=white,draw] (7) at (0,4) {};
\node[fill=white,draw=none, label=center:\scriptsize$\times$, draw] (8) at (1,3) {};
\node[fill=black] (9) at (1,4) {6};
\node[fill=white,draw] (10) at (2,0) {};
\node[fill=black] (11) at (2,1) {1};
\node[fill=white,draw=none, label=center:\scriptsize$\times$, draw] (12) at (2,2) {};
\node[fill=black] (13) at (2,3) {3};
\node[fill=white,draw=none, label=center:\scriptsize$\times$, draw] (14) at (2,4) {};
\node[fill=white,draw] (15) at (3,0) {};
\node[fill=white,draw=none, label=center:\scriptsize$\times$, draw] (16) at (3,1) {};
\node[fill=black] (17) at (3,2) {2};
\node[fill=white,draw=none, label=center:\scriptsize$\times$, draw] (18) at (3,3) {};
\node[fill=black] (19) at (3,4) {4};
\node[fill=white,draw] (20) at (4,0) {};
\node[fill=black] (21) at (4,1) {5};
\node[fill=white,draw=none, label=center:\scriptsize$\times$, draw] (22) at (4,2) {};
\node[fill=white,draw] (23) at (4,3) {};
\node[fill=white,draw] (24) at (4,4) {};

\draw (0)  to (1);
\draw (0)  to (2);
\draw (1)  to (0);
\draw (1)  to (3);
\draw (1)  to (4);
\draw (2)  to (0);
\draw (2)  to (4);
\draw (2)  to (10);
\draw (2)  to (11);
\draw (3)  to (1);
\draw (3)  to (5);
\draw (3)  to (6);
\draw (4)  to (1);
\draw (4)  to (2);
\draw (4)  to (6);
\draw (4)  to (11);
\draw (4)  to (12);
\draw (5)  to (3);
\draw (5)  to (7);
\draw (5)  to (8);
\draw (5)  to (4);
\draw (6)  to (3);
\draw (6)  to (4);
\draw (6)  to (8);
\draw (6)  to (12);
\draw (7)  to (5);
\draw (7)  to (9);
\draw (8)  to (5);
\draw (8)  to (6);
\draw (8)  to (9);
\draw (8)  to (13);
\draw (9)  to (7);
\draw (9)  to (8);
\draw (9)  to (14);
\draw (9)  to[bend right=15] (2);
\draw (10)  to (2);
\draw (10)  to (11);
\draw (10)  to (15);
\draw (10)  to (1);
\draw (11)  to (4);
\draw (11)  to (10);
\draw (11)  to (12);
\draw (11)  to (16);
\draw (12)  to (6);
\draw (12)  to (11);
\draw (12)  to (13);
\draw (12)  to (17);
\draw (13)  to (8);
\draw (13)  to (12);
\draw (13)  to (14);
\draw (13)  to (18);
\draw (13)  to (2);
\draw (14)  to (9);
\draw (14)  to (13);
\draw (14)  to (19);
\draw (15)  to (10);
\draw (15)  to (16);
\draw (15)  to (20);
\draw (15)  to (11);
\draw (16)  to (11);
\draw (16)  to (15);
\draw (16)  to (17);
\draw (16)  to (21);
\draw (16)  to (12);
\draw (17)  to (12);
\draw (17)  to (16);
\draw (17)  to (18);
\draw (17)  to (22);
\draw (17)  to[bend right=15] (2);
\draw (18)  to (13);
\draw (18)  to (17);
\draw (18)  to (19);
\draw (18)  to (23);
\draw (18)  to (24);
\draw (19)  to (14);
\draw (19)  to (18);
\draw (19)  to (24);
\draw (19)  to (22);
\draw (20)  to (15);
\draw (20)  to (21);
\draw (21)  to (16);
\draw (21)  to (20);
\draw (21)  to (22);
\draw (22)  to (17);
\draw (22)  to (21);
\draw (22)  to (23);
\draw (22)  to (2);
\draw (23)  to (18);
\draw (23)  to (22);
\draw (23)  to (24);
\draw (23)  to (19);
\draw (24)  to (19);
\draw (24)  to (23);
\draw (24)  to[bend left=15] (20);
\end{tikzpicture}
\caption{PAGTC: $f_K(S_{\text{P}})=14.$}
\label{f:max_semi}
\end{subfigure}%
\caption{Maximization of $f_K$, with $K=3$ and $r=7$. The solution set $S$ is depicted with black nodes,
with numbers denoting the selection order. The nodes outside the solution $S$ that have at least $K$ neighbors in $S$ are marked with $\times$.}
\label{f:comparison_grid}
\end{figure}

The PAGTC approach clearly outperforms the greedy algorithm in \emph{one-round} influence, often achieving near-optimal solutions on small graphs while maintaining the same computational cost. The specific realization presented in Figure~\ref{f:max_greedy}, drawn from the set of \texttt{small-world} graphs, highlights a fundamental limitation of the greedy strategy: because complex contagion requires multiple active neighbors to influence a new node, the immediate marginal gains during the initial iterations are inherently zero. Consequently, the greedy algorithm makes arbitrary selections, resulting in a suboptimal seed set. In contrast, the PAGTC approach possesses greater long-term planning capabilities, distributing the nodes more effectively to achieve a higher final influence (Figure~\ref{f:max_semi}).

For the \emph{full} influence problem, PAGTC tends to outperform the greedy algorithm for higher values of~$K$, where the increased problem complexity drastically reduces the effectiveness of the greedy approach. The \texttt{fb-messages} network provides an emblematic example: for $K=4$ and $K=5$, the demanding full-influence greedy algorithm yields no improvement over its one-round counterpart, despite evaluating the entire cascade at every step. Meanwhile, the one-round PAGTC surrogate solutions successfully translate into highly effective seed sets for the full cascade.

\begin{figure}
\centering
\begin{subfigure}{.48\linewidth}
\begin{tikzpicture}
\begin{semilogyaxis}[
    width=\linewidth,
    height=5cm,
    axis lines=left,
    grid=both,
    xlabel={$|V|$},
    ylabel={time (s)},
    y label style={yshift=-5pt},
    ytick distance=10,
    legend style={font=\footnotesize},
    legend pos=south east,
]
\addplot[mplorange, thick, mark=*, mark options={scale=0.6}] table {figures/running_time_full_greedy.txt};
\addlegendentry{Greedy}
\addplot[mplblue, thick, mark=triangle*, mark options={scale=.9}] table {figures/running_time_pagtc_delta.txt};
\addlegendentry{PAGTC}
\end{semilogyaxis}
\end{tikzpicture}
\subcaption{Execution time (in seconds) of full-influence maximization algorithms.}
\label{f:exec_time_comparison}
\end{subfigure}\hfill
\begin{subfigure}{.48\linewidth}
\begin{tikzpicture}
\begin{axis}[
    width=6cm,
    height=5cm,
    grid=both,
    ymin=0.9, ymax=1.9,
    xlabel={$|V|$},
    ylabel={influence ratio},
    legend pos=south east,
    legend style={font=\footnotesize},
    axis lines=left,
]
\addplot[mplgreen, thick, mark=*, mark options={scale=0.6}] table {figures/score_ratio.txt};
\addlegendentry{$f_K^*(S_{\text{{P}}})/f_K^*(S_{\text{{G}}})$}
\end{axis}
\end{tikzpicture}
\subcaption{Influence ratio between $S_{\text{{P}}}$ (the PAGTC solution) and $S{_\text{{G}}}$ (the greedy solution).}
\label{f:comparison_ratio}
\end{subfigure}
\caption{Comparison between the greedy and PAGTC algorithms for full influence maximization on a family of small-world graphs of increasing size, each with an average degree above $7$. The parameter~$K$ is set to $5$, while $r$ corresponds to approximately $10\%$ of the total number of nodes.}
\label{f:greedy_pagtc_comparison}
\end{figure}

In terms of computational complexity, the PAGTC algorithm runs in $\mathcal O(r|E|)$ time, a significant improvement over the $\mathcal O(r|V||E|)$ requirement of the greedy approach. Figure~\ref{f:exec_time_comparison} reports the empirical execution times of both algorithms for the full-influence problem across a family of navigable small-world graphs of increasing size. As shown, the PAGTC approach is more than an order of magnitude faster. Furthermore, despite its lower computational cost, it delivers solutions that are approximately $40\%$ to $60\%$ better in terms of final influence (Figure~\ref{f:comparison_ratio}).

\section{Conclusions and Future Work}
\label{s:conclusions}

We introduced \emph{past-aware game-theoretic centrality} (PAGTC), an extension of game-theoretic centrality that conditions node evaluation on a preselected subset of collaborators. Building upon this concept, we developed a general sequential framework for cardinality-constrained set-function maximization on networks. The resulting heuristic is theoretically grounded in selecting, at each step, the node that maximizes its expected marginal contribution to the final solution set. While our framework encompasses the standard greedy strategy as a special case (Remark~\ref{r:greedy_recover}), the PAGTC strategy explicitly incorporates the target budget, contrasting with greedy's local evaluation over the current partial solution.

To demonstrate the practical viability of our approach, we derived computationally efficient expressions for PAGTC scores across two main problem classes: facility location and complex contagion influence maximization (alongside an application to the densest subgraph problem). These explicit formulas enable our method to run with the same time complexity as standard greedy heuristics.

Our empirical findings highlight a clear distinction depending on the objective function. For facility location, where the standard greedy heuristic already achieves strong approximation guarantees, PAGTC yields a weaker theoretical bound and lacks substantial empirical advantages on real-world networks. It can, however, provide tighter data-dependent \emph{a posteriori} bounds, as well as effectively break initial ties in symmetric topologies. Conversely, the method demonstrates clear improvements for non-submodular objectives related to complex contagion. In one-round influence maximization, it consistently outperforms the greedy baseline. More importantly, when one-round PAGTC solutions serve as candidate seed sets for the full influence problem, they consistently surpass the more computationally intensive full-cascade greedy algorithm, particularly under higher activation thresholds, where contagion dynamics are more constrained.

Several future research directions remain open. One immediate next step is to identify new non-submodular objective functions with efficiently computable PAGTC scores, which would broaden the framework's applicability. From a theoretical perspective, establishing formal approximation guarantees for PAGTC in these non-submodular settings remains an open challenge. Another natural extension involves alternative probability distributions, such as the uniform distribution underlying the Shapley value, to handle scenarios without a fixed selection budget. Furthermore, the PAGTC approach may be adapted to more flexible selection paradigms, such as dynamic schemes that alternate between adding and removing nodes during the optimization process. Finally, it would be worthwhile to investigate whether established algorithmic accelerations from the greedy optimization literature can be integrated into the PAGTC framework.

\section*{Data Statement}
The relevant Python code used for the experiments presented here is available in a GitHub repository \texttt{\href{https://github.com/francesco-zigliotto/pagtc}{francesco-zigliotto/pagtc}}.

\section*{Acknowledgments}

The author gratefully acknowledges Professors M.~Benzi, F.~Durastante, and A.~Caraceni for their insightful discussions, encouragement, and invaluable support during the preparation and review of this manuscript. The author is a member of the Gruppo Nazionale di Calcolo Scientifico (GNCS) of the Istituto Nazionale di Alta Matematica (INdAM)

\section*{\refname}
\printbibliography[heading=none]

\end{document}